\documentclass[review]{elsarticle}

\usepackage{lineno,hyperref}
\usepackage{booktabs}
\usepackage{amsmath}
\usepackage{colortbl}
\modulolinenumbers[5]

\journal{Atmospheric Research}

\begin{document}

\begin{frontmatter}

\title{The criterion for infinite positron feedback in dynamics of relativistic runaway electron avalanches}

\author[1,2]{E. Stadnichuk \corref{mycorrespondingauthor}}
\cortext[mycorrespondingauthor]{Egor Stadnichuk}
\ead{yegor.stadnichuk@phystech.edu}
\author[3]{E. Svechnikova}
\address[1]{Moscow Institute of Physics and Technology - 1 “A” Kerchenskaya st., Moscow, 117303, Russian Federation}
\address[2]{HSE University  - 20 Myasnitskaya ulitsa, Moscow 101000 Russia}
\address[3]{Institute of Applied Physics of RAS - 46 Ul'yanov str., 603950, Nizhny Novgorod, Russia}

\begin{abstract}
Relativistic runaway electron avalanches (RREA) accelerated by thunderstorm large-scale electric fields are one of the sources of atmospheric gamma radiation. In strong electric fields, RREAs can multiply by the relativistic feedback. Infinite relativistic feedback makes avalanches self-sustainable and hypothetically can cause a terrestrial gamma-ray flash (TGF). This paper introduces a kinetic approach to study the relativistic feedback caused by positrons since positron feedback dominates for the directly observed electric field strengths. With this approach, the criterion for infinite positron feedback within thunderstorms is derived. Discovered criterion allows obtaining the thunderstorm electric field parameters required for infinite positron feedback for any altitude. The possibility of derived thunderstorm conditions is discussed.
\end{abstract}


\end{frontmatter}


\section*{Highlights}
\begin{itemize}
    \item Kinetic formalism to study the relativistic feedback is developed
    \item Analytical expression for the criterion of infinite positron feedback in dynamics of relativistic runaway electron avalanches is obtained
    \item Directly observed parameters of thunderstorm electric field do not achieve predicted infinite positron feedback conditions
\end{itemize}

\section{Introduction}

The study of radiation of electrified clouds is a rapidly developing field within atmospheric physics, initiated by the detection of the first terrestrial gamma-ray flashes (TGFs) \cite{Fishman1994} and intensified after the discovery of the phenomenon of thunderstorm ground enhancement TGE \cite{Chilingarian_2011_natural_accelerator}.
The radiation observed as TGFs and TGEs is assumed to be cased by energetic electrons accelerated in the thunderstorm electric field. The details of the mechanism remain to be discovered, while the basic concept is as follows.
Relativistic electrons can obtain more energy from acceleration by the electric field than they in average lose by interactions with air molecules \cite{wilson_1925}. The resulting accelerated movement is called a runaway \cite{Gurevich1992, Dwyer2007}. Runaway electrons can produce new runaway electrons by Moller scattering on air molecules \cite{Dwyer_2003_fundamental_limit}. Multiplication of runaway electrons leads to formation of relativistic runaway electron avalanches (RREAs). Electric field strength necessary for RREA production is called critical electric field and depends on the air density \cite{Dwyer_2003_fundamental_limit}.

Runaway electrons can radiate bremsstrahlung gamma-rays when they interact with air. These gamma-rays are detected as high-energy component of Thunderstorm Ground Enhancement (TGE) \cite{Chilingarian_2011_natural_accelerator, Chilingarian_2020_radon}. RREAs are believed to cause thunderstorm gamma-ray glows \cite{Wada2019}. In addition, RREA bremsstrahlung is considered as one of possible sources of Terrestrial Gamma-ray Flashes (TGFs) \cite{asim_spectrum, 10_month_ASIM, Fermi_2016, Dwyer2012_phenomena}. TGF differs from the other high-energy atmospheric physics phenomena in its short duration and high brightness. For RREAs to cause TGF, a large number of RREAs is required \cite{Dwyer_vs_Gurevich}.

In strong large-scale electric fields, relativistic feedback mechanisms influence RREA dynamics \cite{Dwyer_2003_fundamental_limit, Dwyer2012_phenomena}. Consider a thunderstorm electric field region with above critical electric field strength. Secondary cosmic rays can produce seed runaway electrons at the beginning of the region \cite{Gurevich_2001}. This electron can produce a RREA. This avalanche grows while propagating toward the end of the electric field region and radiates gamma-rays. Part of gamma-rays backscatter to the beginning of the region via the Compton scattering and there produce secondary RREAs. In this way, the primary avalanche reproduces itself by the gamma-ray feedback. Another feedback mechanism is the positron feedback. RREA bremsstrahlung gamma-rays can produce electron-positron pairs at the end of the accelerating region. Positrons are accelerated in the opposite to electrons directions, therefore, they run away and reach the beginning of the region and produce secondary RREAs via the Bhabla scattering \cite{Dwyer2007}. In addition, positrons radiate bremsstrahlung towards the beginning of the electric field region, and the radiated gamma can produce secondary avalanches. The probability for a seed RREA to reproduce itself via relativistic feedback is described with feedback coefficient \cite{Dwyer_2003_fundamental_limit}, which shows the ratio of the fluxes of relativistic particles after feedback to the flux of particles before the feedback. If the feedback coefficient is more than 1, the feedback becomes infinite, which means that avalanches multiply and become self-sustainable, as the number of relativistic particles grows with each RREAs generation. Infinite relativistic feedback hypothetically can cause TGFs \cite{asim_spectrum,Dwyer_vs_Gurevich}. For relatively low electric field values, positron feedback dominates over gamma-ray feedback \cite{Dwyer2007}, which motivates to study the positron feedback mechanism in the first place.

In this paper, an analytical solution for RREAs dynamics with positron feedback is derived. The proposed approach is based on the physics of an individual RREA, which has been thoroughly studied in preceding works. RREAs are commonly studied via numerical calculations \cite{Babich_2020} and Monte Carlo simulations \cite{Dwyer_2012, Lehtinen1999, Skeltved2014, Zelenyi_Dwyer, Khamiton2020}. Analytical solutions for individual RREAs were described in \cite{Lehtinen1999, Gurevich_2001, Babich_2020}. A numerical solution of a kinetic equation considering relativistic feedback has been presented in \cite{Dwyer_2012, Liu2013}. RREAs physics in complex thunderstorm electric filed structures was firstly studied in \cite{reactor,doi:10.1063/1.5130111}. Feedback coefficients and infinite feedback conditions were predicted by Monte Carlo simulations \cite{Dwyer2007, Stadnichuk2019}. \cite{Dwyer2007} proposed that the feedback coefficient is proportional to $exp\left(\frac{L}{\lambda_{RREA}}\right)$, where L --- electric field region length, $\lambda_{RREA}$ --- RREA exponential growth length, which can be used as a tool to extrapolate infinite feedback conditions \cite{reactor}. Nevertheless, an analytical model is required for a complete understanding of relativistic feedback physics.

The important question is whether positron feedback conditions are met in thunderstorms or not. To answer this question the infinite feedback conditions obtained in \cite{Dwyer_2012} for sea-level air density should be accurately rescaled to thunderstorm altitudes. In this paper, a kinetic approach to study positron feedback is developed (Section \ref{section_kinetics}, Positron feedback kinetics). As a result, a formula for positron feedback coefficient is derived. The connection between the obtained feedback coefficient and the feedback coefficient defined by \cite{Dwyer_2003_fundamental_limit} is discussed in Section \ref{section_coefficient}, Positron feedback coefficient. With this formula the conditions required for infinite feedback (electric field strength, electric field region length, thunderstorm altitude) are predicted. The scaling of high-energy atmospheric processes is discussed (Section \ref{section_scaling}, Scaling of high-energy process lengths). Infinite feedback conditions obtained by the analytical formula derived in this paper are verified with GEANT4 simulations (Section \ref{section_Geant4}, Positron feedback calculation with GEANT4). The possibility of the infinite positron feedback conditions in thunderstorms in discussed in Section \ref{section_discussion}, Discussion.

\section{Positron feedback kinetics}
\label{section_kinetics}

RREAs dynamics with positron feedback within a region with uniform critical electric field can be described step by step via a relatively simple one-dimensional kinetic approach. A a RREA-accelerating region within a thunderstorm is called a cell\cite{reactor}. If a RREA starts at the point $z_0$ from a single seed relativistic electron then its dynamics can be described, on average, as follows \cite{Gurevich1992, Dwyer2007, Babich_2020}:

\begin{linenomath*}
\begin{equation}
N_{RREA}(z, z_0) = e^{\frac{z - z_0}{\lambda_{RREA}}}
\label{rrea_growth}
\end{equation}
\end{linenomath*}

\begin{figure}

    \centering
    \includegraphics[width=1.0\linewidth]{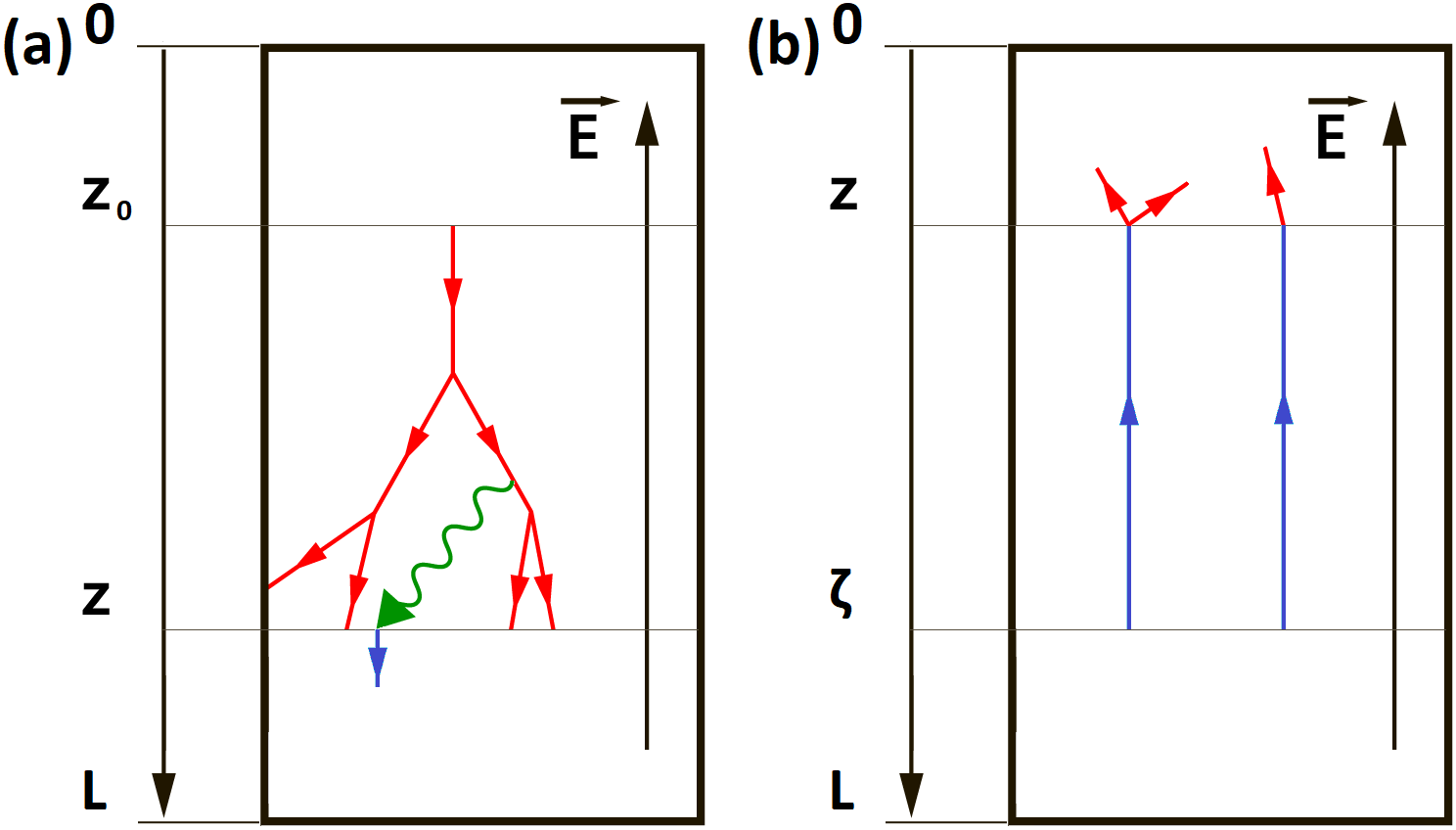}
    \caption{(a) The illustration of the coordinate system used in the kinetic calculation. Red lines --- runaway electrons, green line --- bremsstrahlung gamma-ray. Blue line --- positron generated by the gamma-ray via electron-positron pair production at the point z.
    (b) The illustration of the coordinate system used to calculate secondary avalanches generated by positrons. Blue lines --- positrons, red lines --- secondary runaway electrons produced by these positrons at the point z. Only positrons generated at points $\zeta$ below z coordinate contribute to secondary RREA production at the point z.}
    \label{illustration_1}

\end{figure}

Here, $N_{RREA}(z, z_0)$ --- number of runaway electrons at the point $z$ within a RREA starting from the point $z_0$ (Figure \ref{illustration_1} (a)). $\lambda_{RREA}$ --- RREA e-folding length, which can be calculated, for example, via the empirical formula \cite{Dwyer2007, Babich_2020}:

\begin{linenomath*}
\begin{equation}
\lambda_{RREA} = \frac{U_{RREA}}{E - E_{crit,0} \cdot \frac{\rho}{\rho_0}} = \frac{7300 ~\left[kV\right]}{E - 276 \left[\frac{kV}{m}\right] \cdot \frac{\rho}{\rho_0}}
\label{Dwyer_e-folding}
\end{equation}
\end{linenomath*}

Here $E$ --- uniform electric field value in kV per m, $E_{crit,0}$ --- critical electric field for sea-level air density, $\frac{\rho}{\rho_0}$ --- relative air density for the given thunderstorm environment, $\rho_0$ --- atmosphere air density at the sea level. Runaway electrons naturally produce gamma-rays via bremsstrahlung. Let $\lambda_{\gamma}$ be mean path of a runaway electron for radiation of a gamma-ray photon with energy more than, approximately, 1.022 MeV; this energy is necessary for electron-positron pair production. Let $\lambda_{-}$ be these gamma-rays attenuation length: the mean path before a gamma-ray loses its energy so that it cannot produce positrons. Therefore, the number of energetic gamma-ray photons produced by a RREA can be described with the following equation:

\begin{linenomath*}
\begin{equation}
dN_{\gamma}(z, z_0) = N_{RREA}(z, z_0)\frac{dz}{\lambda_{\gamma}} - N_{\gamma}(z, z_0) \frac{dz}{\lambda_-}
\label{gamma_equation}
\end{equation}
\end{linenomath*}

The first term describes the production of gamma-ray photons by runaway electrons and the second term describes the decrease in the number of gamma-ray photons due to interaction with air. The boundary condition for this equation is $N_{\gamma}(z_0, z_0) = 0$, which leads to the following solution:

\begin{linenomath*}
\begin{equation}
N_{\gamma}(z, z_0) = \frac{\lambda_{RREA} \lambda_{-}}{\lambda_{\gamma}(\lambda_- + \lambda_{RREA})} \cdot \left(e^{\frac{z - z_0}{\lambda_{RREA}}} - e^{-\frac{z - z_0}{\lambda_-}}\right)
\end{equation}
\end{linenomath*}

Let $\lambda_{+}$ be gamma-ray photon mean path for positron production. Thus, positrons are produced by a RREA in the following way:

\begin{linenomath*}
\begin{equation}
\frac{d N_{+}(z, z_0)}{d z} = N_{\gamma}(z, z_0) \frac{1}{\lambda_{+}}
\end{equation}
\end{linenomath*}

To provide relativistic feedback, a positron must reverse in the direction opposite to the RREA propagation direction and run away towards the beginning of the cell. Not all produced positrons reverse, therefore, there is a certain probability of positron reversal \cite{Dwyer_2012}. Let this probability be equal $P_+$. Positrons, propagating to the beginning of the electric field region, on the one hand, annihilate with the mean length $\lambda_{x}$ and, on the other hand, produce secondary runaway electrons with mean path $\lambda_2$. Similar to positrons, these runaway electrons can reverse and produce secondary RREAs \cite{Dwyer_2012, Stadnichuk2019}. Let the probability of secondary RREA development by a secondary runaway electron be equal $P_-$. According to \cite{Dwyer_2012}, reversal probabilities can be found from the following empirical formulas (for $E > \frac{\rho}{\rho_0} \cdot 284 \left[\frac{kV}{m}\right]$):

\begin{linenomath*}
\begin{equation}
P_+ = 0.84 \cdot \left( 1 - exp\left( \frac{-E  + \frac{\rho}{\rho_0} \cdot 150 \left[\frac{kV}{m}\right]}{\frac{\rho}{\rho_0} \cdot 400\left[\frac{kV}{m}\right]} \right) \right)\left( 1 - exp\left( \frac{-E + \frac{\rho}{\rho_0} \cdot 276 \left[\frac{kV}{m}\right]}{\frac{\rho}{\rho_0} \cdot 55\left[\frac{kV}{m}\right]} \right) \right)
\label{reversal_positron}
\end{equation}
\end{linenomath*}

\begin{linenomath*}
\begin{equation}
\begin{split}
P_- = 0.582 \cdot \left( 1 - exp\left( \frac{-E + \frac{\rho}{\rho_0} \cdot 235 \left[\frac{kV}{m}\right]}{\frac{\rho}{\rho_0} \cdot 310 \left[\frac{kV}{m}\right]} \right) \right) + S \left( E - \frac{\rho}{\rho_0} \cdot 1000 \left[\frac{kV}{m}\right] \right) \cdot \\
\cdot 0.268 \cdot \left( \frac{E - \frac{\rho}{\rho_0} \cdot 1000 \left[\frac{kV}{m}\right]}{\frac{\rho}{\rho_0} \cdot 2000\left[\frac{kV}{m}\right]} \right)
\end{split}
\label{reversal_electron}
\end{equation}
\end{linenomath*}

Here, S is the Heaviside step function. Therefore, the number of secondary RREAs produced on the interval $(z, z + dz)$ via positron feedback by primary RREA started at the point $z_0$ (Figure \ref{illustration_1} (b)) is:

\begin{linenomath*}
\begin{equation}
df_2(z, z_0) = dz \cdot \frac{P_+ P_-}{\lambda_{2}} \cdot \int_z^L d\zeta \frac{\partial N_+(\zeta, z_0)}{\partial \zeta} e^{-\frac{\zeta - z}{\lambda_{x}}}
\label{f_2_not_integrated}
\end{equation}
\end{linenomath*}

In this equation (Formula \ref{f_2_not_integrated}), the integral describes number of positrons that reach the interval $(z, z + dz)$. Assuming the electric field value and air density are constant, the integration leads to the following equation:

\begin{linenomath*}
\begin{equation}
\begin{split}
\frac{df_2(z, z_0)}{dz} = \frac{P_+ P_- \lambda_{RREA} \lambda_{-}}{\lambda_{2} \lambda_{+} \lambda_{\gamma}(\lambda_- + \lambda_{RREA})} \bigg[ \frac{\lambda_{RREA}\lambda_{x}}{\lambda_{x} - \lambda_{RREA}} e^{-\frac{z_0}{\lambda_{RREA}}} e^{\frac{z}{\lambda_{x}}} \cdot \\
\cdot \left( e^{\frac{L(\lambda_{x} - \lambda_{RREA})}{\lambda_{x}\lambda_{RREA}}} - e^{\frac{z(\lambda_{x} - \lambda_{RREA})}{\lambda_{x}\lambda_{RREA}}} \right)  - \frac{\lambda_{-}\lambda_x}{\lambda_- - \lambda_x} e^{-\frac{L - z_0}{\lambda_-}} \left(1 - e^{-\frac{L - z}{\lambda_{x}}} e^{\frac{L - z}{\lambda_{-}}} \right) \bigg]
\end{split}
\label{f_2}
\end{equation}
\end{linenomath*}

Function from equation \ref{f_2} describes the number of secondary RREAs produced on the interval $(z, z + dz)$ via positron feedback by a primary RREA, which starts at the point $z_0$, and the origin of the primary avalanche does not matter: it can be a primary RREA produced by secondary cosmic rays or an avalanche produced by relativistic feedback. Therefore, RREAs of the third generation distribution can be obtained with equation \ref{f_2} in the following way. Let the primary (seed) avalanche start at the point $z_0 = 0$. Thus, the secondary RREAs distribution will be $f_2(z, 0)$. That means that within the interval $(\zeta, \zeta + d\zeta)$ the following number of secondary RREAs start their development: $\frac{df_2(\zeta, 0)}{d\zeta}d\zeta$. Each of this avalanches will produce third generation RREAs with the distribution from equation \ref{f_2}: $\frac{df_2(z, \zeta)}{dz}$. To obtain the total third generation avalanches distribution, contribution from all secondary avalanches from the section $[0, L]$ should be summed up, where L is the cell length:

\begin{linenomath*}
\begin{equation}
\frac{d f_3(z, 0)}{d z} = \int^L_0 d \zeta \frac{\partial f_2(z, \zeta)}{\partial z} \frac{\partial f_{2}(\zeta, 0)}{\partial \zeta}
\label{tertiary_avalanches}
\end{equation}
\end{linenomath*}

Similarly to the third RREAs generation, RREAs starting points distribution in $\left( i+1 \right)^{th}$ generation can be obratained from the $i^{th}$ generation as follows:

\begin{linenomath*}
\begin{equation}
\frac{d f_{i+1}(z, 0)}{d z} = \int^L_0 d \zeta \frac{\partial f_2(z, \zeta)}{\partial z} \frac{\partial f_{i}(\zeta, 0)}{\partial \zeta}
\label{feedback_operator}
\end{equation}
\end{linenomath*}

Therefore, RREAs distribution in the next generation can be found as a feedback operator $\hat{F}$ action on the previous generation distribution $\frac{\partial f_i(z, 0)}{\partial z}$ with an operator core $F(z, \zeta)=\frac{\partial f_2(z, \zeta)}{\partial z}$. Thus, the study of positron feedback is reduced to studying the properties of the feedback operator. This approach can be useful, in particular, for relativistic feedback study in variable along the axis $z$ electric field  and air density. Moreover, if a time dependency is added to the operator then it can be used to obtain more accurate relativistic feedback source function for RREA diffusion equation in \cite{Dwyer_2012, Liu2013}. In this paper, electric field and air density are assumed to be constant, moreover, assuming $L \gg \lambda_{RREA}$ and $\lambda_{x} \gg \lambda_{RREA}$, the equation \ref{f_2} simplifies:

\begin{linenomath*}
\begin{equation}
\begin{split}
\frac{df_2(z, z_0)}{dz} \approx \frac{P_+ P_- \lambda_{RREA} \lambda_{-}}{\lambda_{2} \lambda_{+} \lambda_{\gamma}(\lambda_- + \lambda_{RREA})} \cdot \frac{\lambda_{RREA}\lambda_{x}}{\lambda_{x} - \lambda_{RREA}} \cdot \\
\cdot e^{-\frac{z_0}{\lambda_{RREA}}} e^{\frac{z}{\lambda_{x}}} \left( e^{\frac{L(\lambda_{x} - \lambda_{RREA})}{\lambda_{x}\lambda_{RREA}}} - e^{\frac{z(\lambda_{x} - \lambda_{RREA})}{\lambda_{x}\lambda_{RREA}}} \right)
\end{split}
\label{f_2_simple}
\end{equation}
\end{linenomath*}

Therefore, if the primary RREAs starting point distribution is $\propto \delta(z - z_0)$, $z_0 = 0$, then the secondary avalanches distribution is:

\begin{linenomath*}
\begin{equation}
\begin{split}
    \frac{df_2(z, 0)}{dz} \approx \frac{P_+ P_- \lambda_{RREA} \lambda_{-}}{\lambda_{2} \lambda_{+} \lambda_{\gamma}(\lambda_- + \lambda_{RREA})} \cdot \frac{\lambda_{RREA}\lambda_{x}}{\lambda_{x} - \lambda_{RREA}} e^{\frac{z}{\lambda_{x}}} \cdot \\
    \cdot \left( e^{\frac{L(\lambda_{x} - \lambda_{RREA})}{\lambda_{x}\lambda_{RREA}}} - e^{\frac{z(\lambda_{x} - \lambda_{RREA})}{\lambda_{x}\lambda_{RREA}}} \right) = \beta \cdot e^{\frac{z}{\lambda_{x}}} \left( e^{\frac{L(\lambda_{x} - \lambda_{RREA})}{\lambda_{x}\lambda_{RREA}}} - e^{\frac{z(\lambda_{x} - \lambda_{RREA})}{\lambda_{x}\lambda_{RREA}}} \right)
\end{split}
\end{equation}
\end{linenomath*}

\begin{linenomath*}
\begin{equation}
 \beta = \frac{P_+ P_- \lambda_{RREA} \lambda_{-}}{\lambda_{2} \lambda_{+} \lambda_{\gamma}(\lambda_- + \lambda_{RREA})} \cdot \frac{\lambda_{RREA}\lambda_{x}}{\lambda_{x} - \lambda_{RREA}}
 \label{beta}
\end{equation}
\end{linenomath*}

For the third feedback generation:

\begin{linenomath*}
\begin{equation}
    \begin{split}
    \frac{d f_3(z, 0)}{d z} = \beta^2 \frac{\lambda_{RREA} \lambda_{x}}{\lambda_{x} - \lambda_{RREA}} \left( e^{\frac{L(\lambda_{x} - \lambda_{RREA})}{\lambda_x\lambda_{RREA}}} - 1 - \frac{L(\lambda_{x} - \lambda_{RREA})}{\lambda_x\lambda_{RREA}} \right) \cdot \\
    \cdot e^{\frac{z}{\lambda_{x}}} \left( e^{\frac{L(\lambda_{x} - \lambda_{RREA})}{\lambda_{x}\lambda_{RREA}}} - e^{\frac{z(\lambda_{x} - \lambda_{RREA})}{\lambda_{x}\lambda_{RREA}}} \right) = \Gamma \cdot \frac{d f_2(z, 0)}{d z}
    \end{split}
\end{equation}
\end{linenomath*}

$\Gamma$ does not depend on $z$ (the meaning of this value will be discussed below). Further, the following distribution for the $i^{th}$ generation is obtained by induction:

\begin{linenomath*}
\begin{equation}
\frac{df_{i}(z, 0)}{dz} = \Gamma^{i - 2} \cdot \frac{df_2(z, 0)}{dz}
\label{feedback_generations}
\end{equation}
\end{linenomath*}

\begin{figure}
    \centering
    \includegraphics[width=1.0\linewidth]{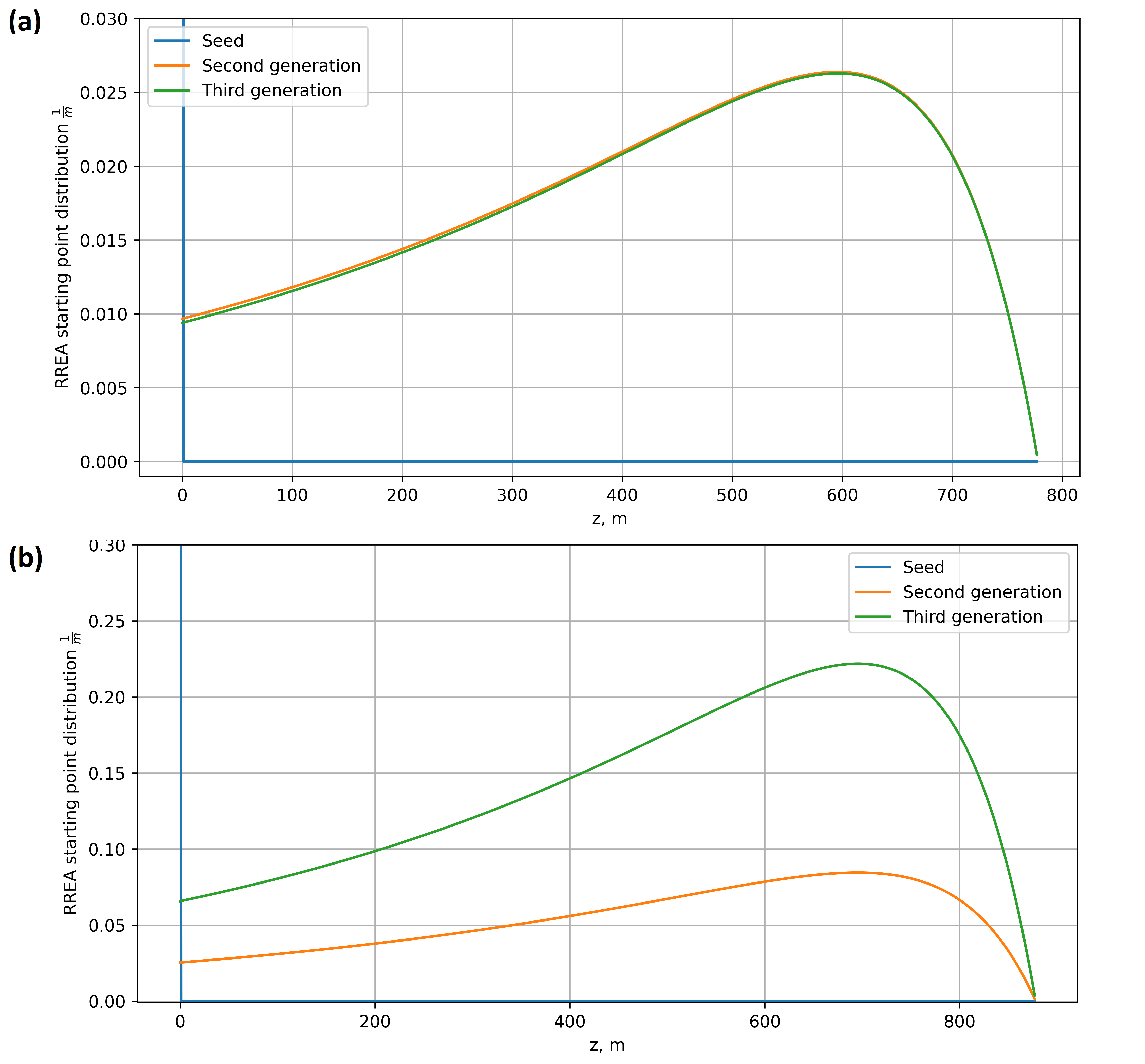}
    \caption{RREA starting point distributions. Distributions calculated with Formulas \ref{f_2}, \ref{tertiary_avalanches}, all terms within Formula \ref{f_2} were considered. High-energy process lengths were retrieved from Section \ref{section_scaling} for altitude 10 km, electric field strength $E = 200$ $\frac{kV}{m}$.
    (a) Positron feedback coefficient $\Gamma \approx 1$ (electric field region length $L = 778$ m). RREA starting point distributions remain the same for all feedback generations starting from the second. Also, it can be seen that the second term in Formula \ref{f_2} does not impact feedback coefficient as it was assumed in Formula \ref{f_2_simple}.
    (b) Positron feedback coefficient $\Gamma = 2.6$ (cell length $L = 878$ m). Distribution shape remains the same for RREAs generation number $\geq 2$, while number of avalanches grows by $\Gamma$ with each generation.
    Number of avalanches can be found by integrating the distribution; number of runaway electrons can be found by integrating the distribution multiplied by $exp\left( \frac{L - z}{\lambda_{RREA}} \right)$ (similar to Formula \ref{electrons_distribution}).}
    \label{RREA_distributions}
\end{figure}

Consequently, the following result is obtained. The RREAs distribution changes from generation to generation only by multiplication by a constant value $\Gamma$ (Figure \ref{RREA_distributions}). $\Gamma$ is a number that shows how many times the number of  relativistic particles changes with the avalanches generation number. In other words, $\Gamma$ --- positron feedback coefficient:

\begin{linenomath*}
\begin{equation}
    \begin{split}
    \Gamma = \frac{P_+ P_- \lambda_{RREA} \lambda_{-}}{(\lambda_- + \lambda_{RREA}) \lambda_2 \lambda_{\gamma} \lambda_{+}} \left(\frac{\lambda_{RREA} \lambda_{x}}{\lambda_{x} - \lambda_{RREA}}\right)^2 \cdot \\
    \cdot \left( e^{\frac{L(\lambda_{x} - \lambda_{RREA})}{\lambda_x\lambda_{RREA}}} - 1 - \frac{L(\lambda_{x} - \lambda_{RREA})}{\lambda_x\lambda_{RREA}} \right)
    \end{split}
\label{feedback_coefficient}
\end{equation}
\end{linenomath*}

Formula \ref{feedback_coefficient} shows that positron feedback coefficient is approximately proportional to $exp\left( \frac{L}{\lambda_{RREA}} \right)$, as it was shown in \cite{Dwyer2007}. It should be noted that the positron feedback coefficient depends only on properties of thunderstorm environment: electric field value, air density and cell length. Similarly to \cite{Dwyer_2003_fundamental_limit, Dwyer2007}, if $\Gamma \geq 1$ then the relativistic feedback becomes infinite (equation \ref{feedback_generations}) and the thunderstorm can possibly produce a TGF, in other cases RREAs decay without external sources of seed particles. Still, if $\Gamma > 0$ then RREAs are more intense than without relativistic feedback by the factor $\frac{1}{1 - \Gamma}$, which can be simply obtained from the geometrical progression (\ref{appendix_b}). Therefore, the criterion for infinite positron feedback in the dynamics of RREAs is as follows:

\begin{linenomath*}
\begin{equation}
    \begin{split}
    \frac{P_+ P_- \lambda_{RREA} \lambda_{-}}{(\lambda_- + \lambda_{RREA}) \lambda_2 \lambda_{\gamma} \lambda_{+}} \left(\frac{\lambda_{RREA} \lambda_{x}}{\lambda_{x} - \lambda_{RREA}}\right)^2 \cdot \\
    \cdot \left( e^{\frac{L(\lambda_{x} - \lambda_{RREA})}{\lambda_x\lambda_{RREA}}} - 1 - \frac{L(\lambda_{x} - \lambda_{RREA})}{\lambda_x\lambda_{RREA}} \right) \geq 1
    \end{split}
\label{feedback_infinite_criterion}
\end{equation}
\end{linenomath*}

\section{Positron feedback coefficient}
\label{section_coefficient}

In the original paper \cite{Dwyer_2003_fundamental_limit}, feedback coefficient is calculated as follows. Avalanches are considered secondary if they are generated by electrons that are generated in the first half of the cell with an upward momentum (that is, the projection of the momentum onto the direction of the electric field is positive). Thus, all secondary avalanches in the first half of the cell are strictly taken into account, since such electrons are generated only by feedback particles flying upward. The feedback coefficient is defined as the ratio of the number of runaway electrons in the secondary avalanche that cross the plane located in the middle of the electric field region to the number of runaway electrons of the primary avalanche that cross this plane. According to Formulas \ref{f_2_simple}, \ref{beta}, the number of secondary avalanches generated by the primary avalanche, starting at point $z_0$, on the segment $[z, z + dz]$ is equal to:

\begin{linenomath*}
\begin{equation}
    \begin{split}
    \frac{\partial f_2(z, z_0)}{\partial z} dz \approx \frac{P_+ P_- \lambda_{RREA} \lambda_{-}}{\lambda_{2} \lambda_{+} \lambda_{\gamma}(\lambda_- + \lambda_{RREA})} \cdot \frac{\lambda_{RREA}\lambda_{x}}{\lambda_{x} - \lambda_{RREA}} e^{-\frac{z_0}{\lambda_{RREA}}} e^{\frac{z}{\lambda_{x}}} \cdot \\
    \cdot \left( e^{\frac{L(\lambda_{x} - \lambda_{RREA})}{\lambda_{x}\lambda_{RREA}}} - e^{\frac{z(\lambda_{x} - \lambda_{RREA})}{\lambda_{x}\lambda_{RREA}}} \right) dz = \beta \cdot e^{-\frac{z_0}{\lambda_{RREA}}} e^{\frac{z}{\lambda_{x}}} \cdot \\
    \cdot \left( e^{\frac{L(\lambda_{x} - \lambda_{RREA})}{\lambda_{x}\lambda_{RREA}}} - e^{\frac{z(\lambda_{x} - \lambda_{RREA})}{\lambda_{x}\lambda_{RREA}}} \right) dz
    \end{split}
\end{equation}
\end{linenomath*}

All secondary avalanches generated on the segment $[0, \frac{L}{2}]$ will generate the following number of runaway electrons on the middle plane:

\begin{linenomath*}
\begin{equation}
    N_2 \left(\frac{L}{2}, z_0\right) = \int_0^\frac{L}{2} dz \frac{\partial f_2(z, z_0)}{ \partial z} e^{\frac{\frac{L}{2} - z}{\lambda_{RREA}}}
\label{electrons_distribution}
\end{equation}
\end{linenomath*}

Consequently, for a primary avalanche starting at the beginning of the electric field region:

\begin{linenomath*}
\begin{equation}
    \begin{split}
    N_2\left(\frac{L}{2}, 0\right) = e^{\frac{L}{2\lambda_{RREA}}} \beta \frac{\lambda_{RREA}\lambda_x}{\lambda_{x} - \lambda_{RREA}} e^{\frac{L(\lambda_{x} - \lambda_{RREA})}{\lambda_{RREA}\lambda_{x}}}\left( 1 - e^{\frac{-L(\lambda_{x} - \lambda_{RREA})}{2\lambda_{RREA}\lambda_{x}}}\right) - e^{\frac{L}{2\lambda_{RREA}}} \beta \frac{L}{2}
    \end{split}
\end{equation}
\end{linenomath*}

According to the definition \cite{Dwyer_2003_fundamental_limit}:

\begin{linenomath*}
\begin{equation}
    \gamma = \frac{N_2\left(\frac{L}{2}, 0\right)}{N_1\left(\frac{L}{2}, 0\right)}
\end{equation}
\end{linenomath*}

$N_1\left(\frac{L}{2}, 0\right) = e^{\frac{L}{2\lambda_{RREA}}}$, hence, the relationship between the feedback coefficient in \cite{Dwyer_2003_fundamental_limit} modeling and the feedback coefficient presented in the criterion is approximately as follows:

\begin{linenomath*}
\begin{equation}
    \gamma \approx \Gamma \cdot \left( 1 - e^{\frac{-L(\lambda_{x} - \lambda_{RREA})}{2\lambda_{RREA}\lambda_{x}}}\right)
    \label{feedback_coefficient_definitions_comparison}
\end{equation}
\end{linenomath*}

It can be seen from formula \ref{feedback_coefficient_definitions_comparison} that $\gamma$ and $\Gamma$ are approximately equal, if $L \gg \lambda_{RREA}$ and $\lambda_{x} \gg \lambda_{RREA}$. Therefore, the definition of the feedback coefficient in this paper and the original definition are consistent. The difference between these definitions is caused by RREA distribution differences between the primary generation and secondary generation. However, starting with the second generation the shape of RREA distribution remains the same (Formula \ref{feedback_generations}). Therefore, if the simulation feedback coefficient is defined as $\gamma = \frac{N_{i+1}\left(\frac{L}{2}, 0\right)}{N_{i}\left(\frac{L}{2}, 0\right)}$, where $i \geq 2$ is the feedback generation number, simulation feedback coefficient $\gamma$ will be exactly equal the kinetic feedback coefficient $\Gamma$. Such approach was presented in \cite{Dwyer2007}, it is more precise than the original approach presented in \cite{Dwyer_2003_fundamental_limit}.

In the previous research \cite{Stadnichuk2019}, positron feedback coefficient has been calculated as the mean number of secondary avalanches produced by a single seed electron in the first half of the strong electric field region. From the kinetic theory this number can be found:

\begin{linenomath*}
\begin{equation}
\begin{split}
    \int_0^{\frac{L}{2}} dz \frac{\partial f_2(z, 0)}{\partial z} = \frac{P_+ P_- \lambda_{RREA} \lambda_{-}}{\lambda_{2} \lambda_{+} \lambda_{\gamma}(\lambda_- + \lambda_{RREA})} \cdot \frac{\lambda_{RREA}\lambda_{x}}{\lambda_{x} - \lambda_{RREA}} \cdot \\
    \cdot \left( \lambda_x e^{\frac{L(\lambda_{x} - \lambda_{RREA})}{\lambda_{RREA}\lambda_{x}}} \left( e^{\frac{L}{2\lambda_x}} - 1\right) - \lambda_{RREA} \left( e^{\frac{L}{2\lambda_{RREA}}} - 1\right) \right) \approx \\
    \approx \Gamma \cdot \frac{\lambda_x}{\lambda_{RREA}}\left( e^{\frac{L}{2\lambda_x}} - 1\right) \approx \Gamma \frac{L}{2\lambda_{RREA}}
\end{split}
\end{equation}
\end{linenomath*}

It can be seen that such approach overestimates the feedback coefficient and, therefore, underestimates the conditions required for infinite feedback.

\section{Scaling of high-energy process lengths}
\label{section_scaling}

It has been shown that gamma-ray process lengths do not depend on the electric field value while they depend on the air density hyperbolically, as $\frac{1}{\rho}$, where $\rho$ is the air density \cite{reactor}. This can be generally applied to processes, which are not related to the electric field directly, for example, positron annihilation length. The reason is that for a large range of electric field strengths the spectrum of RREAs remains approximately the same: $\propto exp\left(-\frac{\varepsilon}{7.3[MeV]}\right)$ \cite{Dwyer2012_phenomena,Cramer_2014_analytical_spectrum,Babich_2020}, and, in the first approximation, electric field value influences only number of runaway electrons per avalanche. Thus, mean gamma-ray production length by a single runaway electron depends on air density: the higher air density is, the higher the frequency of interaction of runaway electrons with molecules. Moreover, as RREA spectrum does not depend on electric field strength, bremsstrahlung gamma-rays spectrum is also field independent and approximately is $\propto \frac{1}{\varepsilon}exp\left(-\frac{\varepsilon}{7.3[MeV]}\right)$ \cite{asim_spectrum}. Therefore, the spectrum and interaction lengths for particles produced by RREA gamma-rays will be, in the first approximation, electric field independent. In what follows, the following notation will be used:

\begin{equation}
    n = \frac{\rho}{\rho_0}
    \label{notation_n}
\end{equation}

Here $\rho$ --- air density for the altitude of interest, $\rho_0$ --- sea-level air density. Air density should be calculated in accordance with the standard atmosphere model. Therefore, high-energy process lengths scale as follows. Gamma-ray photons (with energy sufficient for pair production) production by runaway electron length for 10 km altitude were found as approximately 600 m (RREA gamma-rays were simulated with GEANT4 physics list G4EmStandartPhysics$\_$option4 \cite{geant4}, spatial distribution of gamma-rays was fitted with Formula \ref{gamma_equation}). Therefore, the formula is as follows:

\begin{equation}
    \lambda_{\gamma} = \frac{l_{\gamma}}{n} = \frac{202.53~[m]}{n}    
\end{equation}

According to the NIST database \cite{NIST}, electron-positron production length for 10 km altitude has been estimated as 5000 m (for mean RREA bremsstrahlung gamma-rays energy $\approx 4~MeV$ \cite{asim_spectrum}). It should be noted that if this value is estimated not from Monte Carlo simulations but through electron-positron pair production cross-section, then, as gamma-rays spread at an angle to the electric field, for more accuracy, pair production mean free path should be multiplied by the mean cosine of the angle between gamma-ray momentum and electric field: $\lambda_{+} = \lambda_{\gamma \rightarrow e^+e^-} \cdot cos \alpha$. Moreover, $\lambda_+$ should be averaged by the spectrum of RREA bremsstrahlung. Positron production by gamma length can be estimated as follows:

\begin{equation}
    \lambda_{+} = \frac{l_+}{n} = \frac{1687.8~[m]}{n}
\end{equation}

Gamma decay length for 10 km altitude is approximately 1750 m \cite{reactor}. Therefore, the formula is as follows:

\begin{equation}
    \lambda_- = \frac{l_-}{n} = \frac{590.7~[m]}{n}
\end{equation}

Positron annihilation length was estimated as 500 m for 10 km altitude via cross-sections from GEANT4 source code \cite{geant4}. Therefore, the formula for positron annihilation is as follows:

\begin{equation}
    \lambda_x = \frac{l_x}{n} = \frac{168.8~[m]}{n}
\end{equation}

Mean gamma-ray photon path for a runaway electron production for 10 km altitude was estimated as 1050 m \cite{reactor}. Therefore, the formula for it is as follows:

\begin{equation}
    \lambda_{e^-} = \frac{l_{e^-}}{n} =  \frac{354.4~[m]}{n}
    \label{gamma_to_electron}
\end{equation}

Runaway electrons multiplication within a single avalanche cannot be considered without the influence of the electric field. Therefore, RREA growth length should be estimated with the Formula \ref{Dwyer_e-folding}. In addition, the positron and electron reversal probabilities depend both on air density and electric field strength and can be found according to Formulas \ref{reversal_positron}, \ref{reversal_electron}.

RREA reproduction by positron length was estimated for 10 km altitude and 200 $\frac{kV}{m}$ electric field strength as 106.6 m (GEANT4 simulation). If this value is considered to be independent of the electric field strength, the formula is $\lambda_{2} = \frac{36~[m]}{n}$. Nevertheless, this process is most likely related to the electric field, as positrons run away and are accelerated by the field. Moreover, positrons are in many ways similar to electrons in the sense of interaction with matter. Therefore, RREA reproduction by positron length may be scaled similarly to the Formula \ref{Dwyer_e-folding} for RREA growth length:

\begin{equation}
    \lambda_2 = \frac{U_2}{E - E_{crit,0} \cdot n} = \frac{11400~[kV]}{E - 276\left[\frac{kV}{m}\right] \cdot n}
    \label{RREA_production_positrons}
\end{equation}

With the formulas presented above (Formulas \ref{Dwyer_e-folding}, \ref{reversal_positron}, \ref{reversal_electron}, \ref{notation_n}-\ref{RREA_production_positrons}) the positron feedback coefficient (Formula \ref{feedback_coefficient}) can be rewritten in the following empirical form:

\begin{linenomath*}
\begin{equation}
    \begin{split}
        \Gamma_{empirical} = \frac{n^3 \cdot 9500 \left[ \frac{kV^3}{m^3} \right]}{\left(E - 264\left[\frac{kV}{m} \right] n\right) \left( E - 320\left[\frac{kV}{m} \right] n\right)^2} \cdot \Bigg( exp\left( L \frac{E - 320\left[\frac{kV}{m} \right] n}{7300 \left[ keV\right]} \right) - &\\
        - 1 - L \frac{E - 320\left[\frac{kV}{m} \right] n}{7300 \left[ keV\right]} \Bigg) \cdot \Bigg(0.582 \cdot \left( 1 - exp\left( \frac{-E + n \cdot 235 \left[\frac{kV}{m}\right]}{n \cdot 310\left[\frac{kV}{m}\right]} \right) \right) + &\\
        + S \left( E - n \cdot 1000 \left[\frac{kV}{m}\right] \right) \cdot 0.268 \cdot \left( \frac{E - n \cdot 1000 \left[\frac{kV}{m}\right]}{n \cdot 2000\left[\frac{kV}{m}\right]} \right) \Bigg) \cdot &\\
        \cdot \left( 1 - exp\left( \frac{-E  + n \cdot 150 \left[\frac{kV}{m}\right]}{n \cdot 400\left[\frac{kV}{m}\right]} \right) \right)\left( 1 - exp\left( \frac{-E + n \cdot 276 \left[\frac{kV}{m}\right]}{n \cdot 55\left[\frac{kV}{m}\right]} \right) \right)
    \end{split}
\label{empiric_criterion}
\end{equation}
\end{linenomath*}

\begin{figure}

    \centering
    \includegraphics[width=1\linewidth]{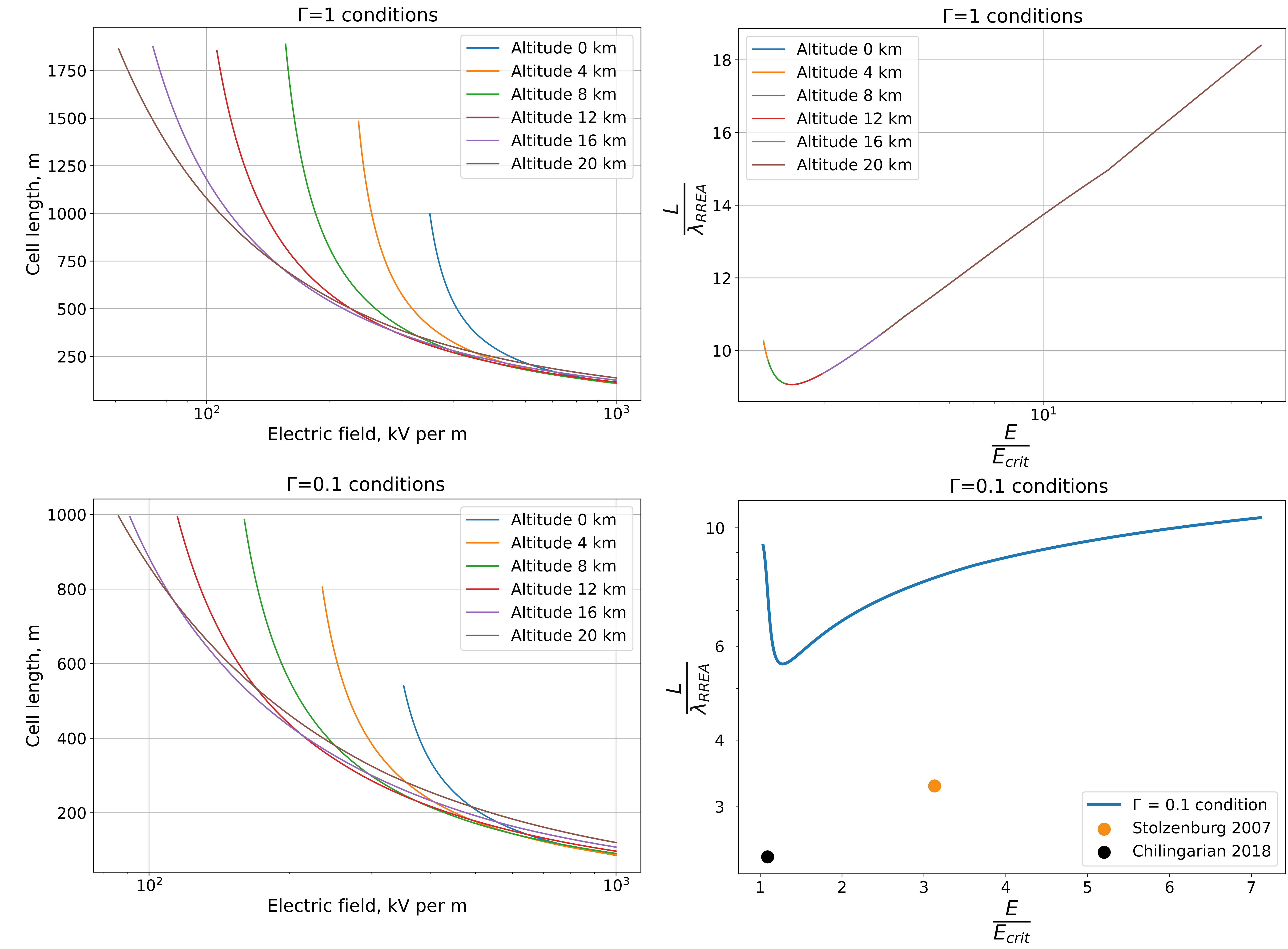}
    \caption{(a) Thunderstorm parameters required for infinite positron feedback (according to Formula \ref{empiric_criterion}). Conditions above the curves are infinite positron feedback conditions with positron feedback coefficient $\Gamma \geq 1$. Conditions below the curves correspond to $\Gamma < 1$, meaning the is no infinite positron feedback. Still, if $\Gamma$ is not much less than 1, positron feedback can influence RREA dynamics (see \ref{appendix_b}).
    (b) Infinite positron feedback conditions presented in the form of the normalized electric field region parameters. X-axis is $\frac{E}{E_{crit}}$, where $E_{crit} \approx 276 n$ --- electric field necessary for RREA development \cite{Dwyer2007}, $n$ is the relative air density (Formula \ref{notation_n}). Y-axis is $\frac{L}{\lambda_{RREA}}$, where $L$ --- electric field region length, $\lambda_{RREA}$ --- avalanche multiplication length (Formula \ref{Dwyer_e-folding}). For all altitudes, the conditions lie on the same curve.
    (c) Positron feedback = 0.1 conditions depending on the electric field region parameters. Under these conditions positron feedback has a noticeable impact on RREAs physics, despite the feedback being not infinite. The conditions are much milder than infinite positron feedback condition on the Figure \ref{conditions_plot} (a).
    (d) Positron feedback = 0.1 conditions depending on the normalized electric field region parameters. The normalization is the same as on the Figure \ref{conditions_plot} (b). Dots mark the conditions described by measured parameters reported in \cite{stolzenburg_2007} and \cite{Chilingarian_2018}. The dots were chosen as the strongest electric fields ever observed in thunderclouds directly and indirectly. It can be seen that these dots do not even reach $\Gamma = 1$ conditions.}
    \label{conditions_plot}


\end{figure}

Figure \ref{conditions_plot} (a) shows necessary for infinite positron feedback thunderstorm electric field parameters calculated with Formula \ref{empiric_criterion} with the necessary condition $\Gamma_{empirical} = 1$. The curves showing infinite positron feedback conditions for different altitudes intersect at high values of the electric field (Figure \ref{conditions_plot} (a)). The reason for it is that for low electric field values, air density makes a significant contribution to the RREA e-folding factor (Formula \ref{Dwyer_e-folding}). The lower the air density is the higher the number of runaway electrons per avalanche. Therefore, for high altitudes, low electric field value (low, but sufficient for RREA development) produces more high-energy particles, including positrons. On the other hand, if the electric field strength is high then the density term in the formula for $\lambda_{RREA}$ (Formula \ref{Dwyer_e-folding}) becomes negligible. That means, that the number of runaway electrons per avalanche ceases to depend on the magnitude of the field. Wherein the electron-positron pair production length increases with altitude. Thus, the positron feedback coefficient decreases with altitude for strong electric field magnitudes. 

Figure \ref{conditions_plot} (b) shows the infinite positron feedback conditions for normalized electric field region parameters. Despite the complexity of the formula for the empirical infinite positron feedback criterion (Formula \ref{empiric_criterion}), the plot turned out to be the same for different atmosphere altitudes. Therefore, the normalization used in Figure \ref{conditions_plot} (b) seems to be useful to describe the conditions required for different kinds of positive feedback \cite{Dwyer2007, reactor}.

It should be noted, that infinite feedback conditions are noticeably dependent on altitude (Figure \ref{conditions_plot} (a)). The reason why Figure \ref{conditions_plot} (b) shows the same infinite feedback conditions for any altitudes can be clearly seen after change of variables in Formula \ref{feedback_coefficient}. Let $\delta_{E} = \frac{E}{E_{crit}} = \frac{E}{E_{crit,0}n}$ and $\delta_{L} = \frac{L}{\lambda_{RREA}}$. Considering the Formulas \ref{Dwyer_e-folding}, \ref{reversal_positron}, \ref{reversal_electron}, \ref{notation_n}-\ref{RREA_production_positrons}), positron feedback coefficient (Formula \ref{feedback_coefficient}) is rewritten in the air density independent form (Formula \ref{feedback_coefficient_normalized}). Reversal probabilities (Formulas \ref{reversal_positron}, \ref{reversal_electron}) only slightly depend on air density when using the variables $\delta_E$ and $\delta_L$. Formula \ref{Dwyer_e-folding} and other empirical formulas presented in this paper have some inaccuracies, therefore, the true shape of the curves in Figure \ref{conditions_plot} (b) may be slightly different.

\begin{linenomath*}
\begin{equation}
    \begin{split}
        \Gamma_{empirical} = \frac{P_+ P_- U_{RREA} l_-E_{crit,0}(\delta_{E} - 1)}{\left(l_-E_{crit,0}(\delta_{E} - 1) + U_{RREA} \right) U_2 l_{\gamma} l_+}\left( \frac{U_{RREA}l_x}{l_x E_{crit,0} (\delta_E - 1) - U_{RREA}}\right)^2 \cdot \\
        \cdot \left(\exp{\left(\delta_{L} \frac{l_x E_{crit,0} (\delta_E - 1) - U_{RREA}}{l_x E_{crit,0} (\delta_E - 1)}\right)} - 1 - \delta_L \frac{l_x E_{crit,0} (\delta_E - 1) - U_{RREA}}{l_x E_{crit,0} (\delta_E - 1)} \right)
    \label{feedback_coefficient_normalized}
    \end{split}
\end{equation}
\end{linenomath*}

The obtained analytical description can be used not only to derive infinite feedback conditions. The conditions when relativistic feedback influences RREA physics can also be predicted. Figures \ref{conditions_plot} (c), \ref{conditions_plot} (d) show the conditions required for the positron feedback coefficient being equal 0.1; in this case positron feedback noticeably influences RREAs but does not make them self-sustainable. The question how a feedback coefficient less than 1 impacts the number of relativistic particles is addressed in \ref{appendix_b}.




\section{Positron feedback simulation with GEANT4}
\label{section_Geant4}

To verify the proposed kinetic theory of relativistic feedback physics, a GEANT4 simulation was carried out \cite{geant4}. In this research, GEANT4 physics list G4EmStandartPhysics$\_$option4 was chosen, because this physics list is acknowledged as the most precise one for high-energy atmospheric physics simulations with GEANT4 \cite{Skeltved2014,Chilingarian_2018,Khamiton2020,Zelenyi_Dwyer}. To separate positron feedback from other physical effects and to divide different feedback generations the simulation was divided in several steps \cite{Stadnichuk2019}. On the first step, seed electrons were launched at the beginning of the electric field region. This electrons form RREAs, which radiate gamma-rays via bremsstrahlung. Energy, position and momentum of positrons generated by these gamma-rays were recorded, the particles themselves were stopped to prevent the relativistic feedback in the first simulation. On the second step, positrons with parameters obtained from the previous simulation were launched. Positrons reversed, traveled to the beginning of the electric field region, and generated runaway electrons. Position, energy, and momentum of these electrons were recorded, electrons themselves were stopped to avoid RREAs formation. In the third step, electrons were launched and second feedback generation particles were registered. Similarly, third feedback generation particles were obtained. According to kinetic theory, RREA spatial distribution stabilizes after second generation of positron feedback. Therefore, feedback coefficient can be calculated via the following formula:

\begin{equation}
    \Gamma_{G4} = \frac{N^3_{total}}{N^2_{total}}
    \label{GEANT4_feedback_coefficient}
\end{equation}

The Formula \ref{GEANT4_feedback_coefficient} is justified similarly to the Section \ref{section_coefficient}, Positron feedback coefficient. The results of the GEANT4 simulations in comparison with Formula \ref{feedback_infinite_criterion} predictions are presented on Table \ref{GEANT4_comparison}. Formula \ref{feedback_infinite_criterion} was proved useful to estimate infinite feedback conditions, as blindfold search for these conditions via GEANT4 simulation takes a considerable amount of calculation time. Nevertheless, the prediction of RREA growth length via Formula \ref{Dwyer_e-folding} is not consistent with GEANT4 simulations calculations of $\lambda_{RREA}$: $\lambda_{RREA} \approx 85~m$ for 10 km altitude and 200 $\frac{kV}{m}$ electric field strength \cite{Khamiton2020} while Formula \ref{Dwyer_e-folding} predicts $\lambda_{RREA} \approx 68~m$. Thus, for 10 km altitude infinite feedback predictions Formula \ref{Dwyer_e-folding} numerator $U_{RREA}$ was changed in order to fit GEANT4 RREA growth length. In this way, the infinite feedback conditions estimation via empirical Formula \ref{empiric_criterion} appeared to be consistent with GEANT4 simulations (Table \ref{GEANT4_comparison}).

\begin{table}[]
\centering
\resizebox{\textwidth}{!}{%
\begin{tabular}{@{}
>{\columncolor[gray]{0.8}}cccccccccccc@{}}
\toprule
E, $\frac{kV}{m}$ & 200   & 210   & 220    & 230   & 240   & 250   & 260    & 270    & 280   & 290    & 300    \\ \midrule
L, m              & 778.0 & 719.5 & 670.25 & 628.2 & 591.8 & 560.0 & 531.85 & 506.85 & 484.4 & 464.15 & 445.75 \\ \midrule
$\Gamma_{G4}$     & 1.16  & 1.21  & 1.06   & 0.97  & 1.05  & 1.10  & 0.98   & 0.92   & 0.89  & 0.93   & 0.94   \\ \bottomrule
\end{tabular}%
}
\caption{The comparison between kinetic theory predictions and GEANT4 simulation results (physics list: G4EmStandartPhysics$\_$option4). The number of seed electrons per simulation is 100 (making feedback coefficient relative error $\approx 10 \%$), the kinetic energy of seed electrons is 3 MeV. E --- electric field strength, L --- RREA accelerating region length necessary for infinite positron feedback predicted via kinetic theory (Formula \ref{feedback_infinite_criterion}), $Gamma_{G4}$ --- feedback coefficient calculated via GEANT4. Within the simulation error, kinetic theory accurately predicts the conditions required for infinite positron feedback.}
\label{GEANT4_comparison}
\end{table}

Points of birth of secondary avalanches and relativistic particles can be played according to the distributions obtained from Section \ref{section_kinetics}, Positron feedback kinetics. Nevertheless, the kinetic theory predicts the average positron feedback behavior. In fact, positron feedback coefficient is a stochastic value. GEANT4 simulations showed that it has a significant fluctuation from RREA to RREA. Feedback coefficient spread is associated with the random nature of relativistic particle production. The feedback coefficient obeys the Poisson distribution, and the dispersion is rather high. Nevertheless, for a huge number of avalanches, the coefficient averages. Considerable number of avalanches is observed during TGFs \cite{Fermi_2016}. In this way, the kinetic study of relativistic feedback physics via averaged values is justified.

\section{Transverse dynamics of the relativistic feedback}
\label{section_transverse}

In this paper, the theoretical model is developed under the assumption that the width of the electric field region is large in comparison with the characteristic scale of a RREA. The width of the electric field region can be taken into account in the following way. Secondary avalanches generated via relativistic feedback are born randomly on the side of the parent avalanche. Therefore, the transverse dynamics of the relativistic feedback can be described as the Brownian motion of secondary avalanches. Consider the mean lateral distance between the parent avalanche and the avalanche produced by the parent via relativistic feedback to be equal to $\lambda_{RF}$, where $RF$ means Relativistic Feedback. Secondary RREAs are generated by gamma-rays and positrons, and the positrons are bound to the electric field line, thus, secondary RREA position is defined by the gamma-ray point of interaction. Therefore, $\lambda_{RF}$ can be estimated via gamma-ray diffusion: $\lambda_{RF} \approx \sqrt{2 D \frac{L}{2 c}} \approx \sqrt{\frac{\lambda L}{6}}$, where $D = \frac{\lambda c}{3}$ --- gamma-ray diffusion coefficient, $\lambda$ --- gamma-ray mean free path length ($\approx 400~m$ for 1 MeV gamma at 10 km altitude \cite{NIST}), L --- electric filed region length (can be defined from Figure \ref{conditions_plot}). Let the average relativistic feedback time be equal $\tau_{RF}$. $\tau_{RF}$ is also the time one one feedback generation. Therefore, the diffusion coefficient of secondary avalanches motion approximately equals $D_{RF} \approx \frac{\lambda_{RF}^2}{\tau_{RF}}$. If the characteristic transverse dimension of the electric field region equals $R$ then the characteristic time required for secondary avalanches to cross the edge of the region can be estimated from the equation $R = \sqrt{2 D_{RF} t_{edge}}$. Thus, the average number of feedback generations required for secondary avalanches to cross the edge of the electric field region can be estimated as $n_{edge} = \frac{t_{edge}}{\tau_{RF}} = \frac{R^2}{2 \lambda_{RF}^2}$. The Brownian motion of secondary avalanches inevitably affects the relativistic feedback physics if the total time of RREA dynamics is comparable with $t_{edge}$. In this case, there is a lateral outflow of RREAs, which leads to the relativistic feedback coefficient decrease and tightens the feedback criterion (Formula \ref{feedback_infinite_criterion}). If $\lambda_{RF} \approx 100~m$, $R \approx 1000~m$, and $\tau_{RF} \approx \frac{L}{c} \approx 1~us$, where L --- electric filed region length and c --- the light speed, the characteristic time required for secondary avalanches to cross the edge of the region is $t_{edge} = n_{edge} \tau_{RF} \approx 50~us$. This time corresponds to the characteristic TGF event time \cite{10_month_ASIM}, thus, in the first approximation, this Brownian motion can be neglected for TGF description via relativistic feedback and the relativistic feedback dynamics can be considered one-dimensional as it is in this paper. Still, for longer events, lateral size of the electric field region should be taken into account.

\section{Discussion}
\label{section_discussion}

The approach of the relativistic feedback study presented in Section Positron feedback kinetics, \ref{section_kinetics}, is based on one-dimensional RREA dynamics analysis via mean high-energy process lengths, while processes occurring in the direction transverse to the electric field are taken into account within the values of mean path lengths of high-energy processes. Moreover, it is shown in Section \ref{section_transverse} that transverse dynamics of RREAs caused by the relativistic feedback is negligible for TGF timescales. The spectrum of a single avalanche stabilizes after several RREA growth lengths \cite{Babich_2020}. Thus, high-energy process macroscopic cross-sections, which are averaged over the spectrum, also stabilize at relatively short distances. This makes RREA growth and gamma-ray and positron production, and gamma-ray and positron spectra stable as well. As RREA processes can be averaged, the whole physics can be described with the presented kinetic approach. 

Gamma-ray feedback can be also described with the presented kinetic approach. To provide a feedback gamma-rays must backscatter. Backscattering can be implemented within the feedback operator as multiplication by the probability $P_{\gamma, back}$. Also, gamma-ray spectrum softening after reversal should be considered within backscattered gamma-ray decay length and mean path for runaway electron production (Formula \ref{gamma_to_electron}). This high-energy process lengths and $P_{\gamma, back}$ can be calculated, for example, with GEANT4 by the study of runaway electron distribution produced by a gamma-ray beam, similarly to the study of gamma-ray process lengths \cite{reactor}. Backscattering will be observed in the runaway electron distribution behind the starting point of the gamma-ray beam.

Feedback operator derived for positron feedback in uniform electric field (Formula \ref{feedback_operator}) is a good tool for studying relativistic feedback and reactor feedback \cite{reactor} in various systems. The feedback operator is built step by step considering all high-energy processes participating in the individual feedback process. Non-uniformities of the electric field strength can be taken into account within the feedback operator (Formula \ref{feedback_operator}) by the consideration of high-energy process lengths dependence on the electric field value and air density, geometrical non-uniformity also can be taken into account. Moreover, the feedback operator can be calculated with Monte Carlo simulations. The advantage of this theoretical approach is that after the operator is built, the feedback mechanism study comes down to eigenvalues and eigenfunctions calculation. In many simple cases, this can be done analytically, as it is for positron feedback in the uniform electric field in this paper.

The formula for the positron feedback coefficient presented in this paper (Formula \ref{feedback_infinite_criterion}) allows studying of relativistic feedback without a thorough Monte Carlo simulation. The analysis of feedback physics via this formula is based on the understanding of individual atmospheric high-energy processes. The empiric formula \ref{empiric_criterion} provides a good estimation for the positron feedback coefficient in the Earth's atmosphere. The necessary conditions for infinite positron feedback can be estimated as $\Gamma_{empirical} = 1$ (Figure \ref{conditions_plot}). Nevertheless, to predict these conditions precisely, a more accurate calculation of high-energy process lengths is required. The reason is that for electric field strengths close to the critical value RREA spectrum softens \cite{Babich_2020}, which has not been taken into account in this paper, while high-energy particles spectrum is the key parameter which influences high-energy process lengths. In addition, positron transport physics should be studied in more detail. Moreover, a further investigation of RREA growth length is required as Formula \ref{Dwyer_e-folding} has discrepancies for high altitudes \cite{Khamiton2020}, which can lead to significant inaccuracies, since this parameter is under the exponent (Formula \ref{rrea_growth}). In addition to a better understanding of the relativistic feedback, a thorough examination of individual atmospheric high-energy processes is required for the study of the reactor feedback in RREA dynamics in complex electric field structures \cite{reactor}.

The most important question is whether infinite positron feedback conditions are met within thunderstorms or not. The most accurate information about the field in a thundercloud can be obtained by balloon and rocket measurements \cite{stolzenburg_2007, marshall1995, Marshall1998_estimates, Stolzenburg_2008_profiles, Dye_2007_fields, Dye_2007_enhancement,  Stolzenburg_1998_precipitation, Stolzenburg_1998_structure, Stolzenburg_2008_profiles, Stolzenburg_2015}.
Balloons rise to a height up to 20 km, the characteristic ascent speed is 10 m/s, the vertical coordinate error is about 10 m. The maximum electric field strength observed in thunderclouds is close to the electric field required for the formation of RREAs \cite{rakov_uman2003} (p.82, s.3.2.4 , table 3.2).  
In addition, information about the electrical structure of a cloud can be obtained by numerical modeling using TGE observations \cite{Chilingarian_2018}. Figure \ref{conditions_plot} shows the analytical estimation of the feedback coefficient depending on the parameters of a strong field region and the maximal parameters of the strong field region found in observations.
Figure \ref{conditions_plot} (d) presents the theoretical estimation of the parameters of high-field region, providing an avalanche multiplication factor of 0.1, as well as the experimental evidence of the electric field above the threshold. Electric field exceeding the value of avalanche formation has been registered using the balloon in the vicinity of a lightning flash (about 100~m from the balloon to the lightning channel) \cite{stolzenburg_2007} and calculated for the TGE-producing cloud discussed in \cite{Chilingarian_2018}. \cite{stolzenburg_2007} reports that the field meter froze at 220 kV/m for 15~s of the balloon ascent at the altitude of about 12 km, which reveals the region of about 160 m height with the electric field exceeding 220 kV/m, corresponding to the feedback coefficient $\Gamma \approx 0.016$ (the cell length can be longer than 160 m, because the balloon likely did not measured the entire region). The modeling based on the measured enhancement of energetic particle flux \cite{Chilingarian_2018} gives the following parameters: electric field exceeding 220 kV/m in the region 160 m height at the altitude of about 12 km). Therefore, it can be concluded that infinite positron feedback is not achievable in the Earth's thunderstorms. Nevertheless, it should be taken into account that balloon measurements provide values of the electric field averaged over vertical distance of about 10~m. On a scale less than 10~m, even larger values of electric field strength can be expected, particularly due to the enhancement of the electric field strength near hydrometeors (\cite{rakov_uman2003}  s.3.2.4). In addition, one should expect that regions with a maximal potential difference have not been registered by experimenters, leaving the opportunity of a higher feedback coefficient.
And finally, it is problematic to measure infinite feedback electric field directly, as the electric field will decrease in approximately 100 us according to the Relativistic Feedback Discharge Model \cite{Dwyer_2012}.

Another interesting question is whether a lightning leader can trigger infinite positron feedback or not \cite{Dwyer2012_phenomena}. Positron feedback coefficient in lightning leader electric field can be estimated with the criterion obtained in this paper. \cite{Moss_2006} estimated the lightning leader streamer zone electric field as 450 kV per m per atm for the positive leader and up to 1250 kV per m per atm for the negative leader. The total potential difference in the streamer zone was estimated as about half of the potential difference inside the thundercloud. In the case of streamer zone potential difference 100 MV, the accelerating region length is about 100 m for the negative leader and 200 m for the positive leader. For normal conditions, the positron feedback coefficient is $\Gamma \approx 0.003$, that is, there is almost no positron feedback influence on RREAs. At high altitudes no more than $\Gamma \approx 0.15$ was obtained. Therefore, there can be some influence of the positron feedback on RREAs accelerating in the lightning leader corona, but infinite positron feedback is not achievable. Gamma-ray feedback hypothetically might play a significant role in the streamer zone electric field under consideration, because gamma-ray feedback is strong in small cells with high electric field values. Nevertheless, it should be noted that the transverse feedback diffusion (Section \ref{section_transverse}) can significantly weaken the relativistic feedback in lightning leader electric field.

The conditions for infinite positron feedback are hard to meet in the Earth's thunderstorms, making infinite positron feedback a less likely scenario for TGFs. However, besides positron feedback, there are other types of positive feedback in RREAs physics. For example, in complex thunderstorm electric structures, reactor feedback appears, and the electric conditions for infinite reactor feedback are significantly softer than for the positron feedback \cite{reactor}. Moreover, in short, and strong electric regions gamma-ray feedback hypothetically might be infinite \cite{Dwyer2007}. Therefore, it is an open question, whether infinite positive feedback can cause TGFs in the Earth's atmosphere.

\section{Conclusions}

A kinetic formalism for studying the relativistic feedback in RREA physics is developed. With the positron feedback operator, the formula for positron feedback coefficient is derived, and the criterion of infinite positron feedback is discovered. Thunderstorm conditions required for infinite feedback defined by the criterion are found to be consistent with GEANT4 simulation predictions. Moreover, the distributions derived from the kinetic model allow the justification of feedback coefficient calculation via modeling methods.

The criterion depends on RREA high-energy process lengths, which should be accurately calculated to make precise predictions on the positron feedback coefficient formula. In this paper, approximate expressions for high-energy atmospheric physics process lengths are introduced. However, a further research is needed to determine the more precise altitude scaling for high-energy atmospheric physics.

The conditions required for infinite positron feedback at different altitudes are calculated. It is shown that directly observed parameters of thunderstorms are not sufficient for infinite positron feedback. Nevertheless, positron feedback coefficient $> 0.01$ is estimated for TGE conditions. This indicates at least a small contribution of positron feedback to RREA physics.

\section*{Acknowledgements}

The work of E. Stadnichuk was supported by the Foundation for the Advancement of Theoretical Physics and Mathematics “BASIS”. The work of E. Svechnikova was supported by a grant from the Government of the Russian Federation (contract no. 075-15-2019-1892).

\bibliography{bib.bib}
\appendix
\section{Distribution of particles generated by RREAs with relativistic feedback}

Distributions of particles generated by RREAs with relativistic feedback can be approximately found as follows. One avalanche, when passing a distance $z$ along a uniform electric field without feedback, gives rise to the following number of particles:

\begin{linenomath*}
\begin{equation}
    N^{particles}(z) = N_{RREA}^{particles} \cdot e^{\frac{z}{\lambda{REEA}}}
\end{equation}
\end{linenomath*}

Here $N_{RREA}^{particles}$ is a constant depending on what kind of particles are born. It can include, in addition to runaway electrons, gamma-ray photons, ions, etc. It is also directly proportional to the number of runaway electrons at a $z = 0$ point. For example, $N_{RREA}^{particles} = N_{RREA}(0)$ for runaway electrons, $N_{RREA}^{particles} = N_{RREA}(0) \frac{\lambda_{RREA} \lambda_{-}}{\lambda_{\gamma}(\lambda_- + \lambda_{RREA})}$ for gamma-rays, $N_{RREA}^{particles} = N_{RREA}(0) \frac{\lambda_{RREA}^2 \lambda_{-}}{\lambda_{\gamma}(\lambda_- + \lambda_{RREA})\lambda_{+}}$ for positrons. According to formulas \ref{f_2_simple}, \ref{feedback_coefficient}, \ref{feedback_generations} distributions of RREAs for every relativistic feedback generation can be found via the following recurrence ratio for $i > 1$:

\begin{linenomath*}
\begin{equation}
    f_{i + 1}(z, 0) = \Gamma \cdot f_i(z, 0)
\end{equation}
\end{linenomath*}

RREAs of the i-th relativistic feedback generation, born on a segment $[z, z + dz]$, form the following number of particles in the entire uniform critical field:

\begin{linenomath*}
\begin{equation}
    dN_{i}^{particles} = dz \cdot \frac{df_i(z, 0)}{dz} \cdot N_{RREA}^{particles}e^{\frac{L - z}{\lambda_{RREA}}}
\end{equation}
\end{linenomath*}

Therefore, the total number of particles born by the i-th RREA generation is:

\begin{linenomath*}
\begin{equation}
    N_{i}^{particles} = \int_0^L dz \cdot \frac{df_i(z, 0)}{dz} \cdot N_{RREA}^{particles}e^{\frac{L - z}{\lambda_{RREA}}}
\end{equation}
\end{linenomath*}

Hence:

\begin{linenomath*}
\begin{equation}
    \begin{split}
    N_{i}^{particles} = \beta \Gamma^{i - 2} N_{RREA}^{particles} \cdot e^{\frac{L}{\lambda_{RREA}}} \cdot \frac{\lambda_{RREA} \lambda_{x}}{\lambda_{x} - \lambda_{RREA}} \cdot \\
    \cdot \left( e^{\frac{L(\lambda_{x} - \lambda_{RREA})}{\lambda_x\lambda_{RREA}}} - 1 - \frac{L(\lambda_{x} - \lambda_{RREA})}{\lambda_x\lambda_{RREA}} \right)
    \end{split}
\end{equation}
\end{linenomath*}

Using formula \ref{feedback_coefficient} the following total number of particles is obtained:

\begin{linenomath*}
\begin{equation}
    N_{i}^{particles} = \Gamma^{i - 1} N_{RREA}^{particles} e^{\frac{L}{\lambda_{RREA}}}
\end{equation}
\end{linenomath*}

The spatial distribution of these particles is as follows:

\begin{linenomath*}
\begin{equation}
    N_{i}^{particles}(z) = \Gamma^{i - 1} N_{RREA}^{particles} e^{\frac{z}{\lambda_{RREA}}}
\end{equation}
\end{linenomath*}

\section{The total number of particles produced by the primary electron at non-infinite feedback conditions}
\label{appendix_b}

Let the primary runaway electron fall into a uniform critical electric field at a point. Without feedback, it will give birth to the following number of particles:

\begin{linenomath*}
\begin{equation}
    N_1^{particles} = N_{RREA}^{particles} \cdot e^{\frac{L}{\lambda{REEA}}}
\end{equation}
\end{linenomath*}

Note that, taking into account the feedback, in the i-th generation of avalanches of runaway electrons, the following number of particles will be generated:

\begin{linenomath*}
\begin{equation}
    N_i^{particles} = \Gamma^{i - 1} \cdot N_{RREA}^{particles} \cdot e^{\frac{L}{\lambda{REEA}}} = \Gamma^{i - 1} \cdot N_1^{particles}
\end{equation}
\end{linenomath*}

Therefore, $\forall i \geq 1$:

\begin{linenomath*}
\begin{equation}
    N_i^{particles} = \Gamma^{i - 1} \cdot N_1^{particles}
\end{equation}
\end{linenomath*}

$\Gamma < 1$, therefore, the total number of particles is found by the formula of an infinitely decreasing geometric progression:

\begin{linenomath*}
\begin{equation}
    N_{\sum}^{particles} = \frac{N_1^{particles}}{1 - \Gamma}
\end{equation}
\end{linenomath*}

\end{document}